\documentclass[aps,prc,twocolumn,floatfix,superscriptaddress,a4paper,
               showpacs,showkeys,nofootinbib]{revtex4}
\usepackage{epsfig}
\usepackage{latexsym}
\usepackage{xspace}
\usepackage{lipsum}
\usepackage{hyperref}
\usepackage[latin2]{inputenc}
\usepackage{indentfirst}
\usepackage{enumerate}
\usepackage{color}

\usepackage{amsmath}
\usepackage{amssymb}
\usepackage[english]{babel}
\usepackage{url}
\newcommand{\eq}[1]{\begin{align} #1 \end{align}}

\begin{document}

\title{Scaled variance, skewness, and kurtosis near the critical point of nuclear matter }
\author{V. Vovchenko}
\affiliation{
Taras Shevchenko National University of Kiev, 03022 Kiev, Ukraine}
\affiliation{
Frankfurt Institute for Advanced Studies, Johann Wolfgang Goethe University,
D-60438 Frankfurt, Germany}
\affiliation{
GSI Helmholtzzentrum f\"ur Schwerionenforschung GmbH, D-64291 Darmstadt, Germany}
\author{D. V. Anchishkin}
\affiliation{
Bogolyubov Institute for Theoretical Physics, 03680 Kiev, Ukraine}
\affiliation{
Taras Shevchenko National University of Kiev, 03022 Kiev, Ukraine}
\affiliation{
Frankfurt Institute for Advanced Studies, Johann Wolfgang Goethe University,
D-60438 Frankfurt, Germany}
\author{M. I. Gorenstein}
\affiliation{
Bogolyubov Institute for Theoretical Physics, 03680 Kiev, Ukraine}
\affiliation{
Frankfurt Institute for Advanced Studies, Johann Wolfgang Goethe University,
D-60438 Frankfurt, Germany}
\author{R. V. Poberezhnyuk}
\affiliation{
Bogolyubov Institute for Theoretical Physics, 03680 Kiev, Ukraine}

\date{\today}

\begin{abstract}
The van der Waals (VDW) equation of state predicts the existence of a first-order
liquid-gas phase transition and contains a critical point.
The VDW equation with Fermi statistics is applied to a description of the
nuclear matter.
The nucleon number fluctuations near the critical point of nuclear matter are studied.
The scaled variance, skewness, and kurtosis  diverge at the critical point.
It is  found that the crossover region of the phase diagram is
characterized by the large values of the scaled variance,
the almost zero skewness, and the significantly negative
kurtosis. The rich structures
of the skewness
and kurtosis are observed in the phase diagram in the wide region around the critical point, namely, they both
may attain large positive or negative values.

\end{abstract}



\maketitle

\section{Introduction}

The first-order liquid-gas phase transition is a well known phenomenon that takes place in
atomic and/or molecular systems, in the system of interacting nucleons (nuclear matter), and
most probably between hadrons and quark-gluon plasma at large baryonic densities.
In all these cases the phase transition line in the plane of temperature $T$ and chemical
potential $\mu$ has an end point. It is called the critical point (CP), and it demonstrates
some universal features typical for the second-order phase transitions, in particular,
anomalously large fluctuations.

The particle number fluctuations
are characterized by the central moments,
$\langle (\Delta N)^2\rangle$,  $\langle (\Delta N)^3\rangle$, and
$\langle (\Delta N)^4\rangle$, etc., where $\langle \ldots\rangle$
denotes the event-by-event averaging and $\Delta N\equiv N-\langle N\rangle$.
The scaled variance $\omega[N]$, skewness $S\sigma$,
and kurtosis $\kappa\sigma^2$ defined as the following combinations of the central moments,
\eq{
\omega[N] & =  \frac{\langle (\Delta N)^2\rangle}{\langle N\rangle}~, 
\nonumber\\
S\sigma & =  \frac{\langle (\Delta N)^3\rangle}{\langle (\Delta N)^2\rangle}~,
\label{N}\\
\kappa\sigma^2 & =  \frac{\langle (\Delta N)^4\rangle - 3\langle (\Delta N)^2\rangle^2}{\langle (\Delta N)^2\rangle}~,
\nonumber
}
are the well known size-independent
(intensive) measures of particle number fluctuations.

In the grand canonical ensemble (GCE) the pressure $p$ plays the role of the thermodynamical
potential, and its natural variables are temperature $T$ and chemical potential $\mu$.
The particle number fluctuations can be characterized by the following dimensionless
cumulants ($n=1,2,\ldots$),
\eq{\label{kn}
k_n = \frac{\partial^n (p/T^4)}{\partial(\mu/T)^n}~.
}
The fluctuation measures in 
Eq.~\eqref{N}
can be then presented as the following:
\eq{\label{osk}
\omega[N]
=\frac{k_2}{k_1}~,~~~~~
S\sigma=\frac{k_3}{k_2}~,~~~~~~ \kappa \sigma^2=\frac{k_4}{k_2}~.
}

The study of event-by-event  fluctuations in high-energy nucleus-nucleus
collisions opens new possibilities to investigate properties of strongly
interacting matter
(see, e.g., Refs.~\cite{Koch:2008ia} and \cite{G-2015} and references therein),
and the experimental search for the chiral CP is now in progress (see, e.g.,
Refs.~\cite{GGS-2014,G-S,c-QCD} and references therein).
The fluctuation signals of the QCD CP were discussed in Refs.~\cite{fluc1,fluc2,fluc3}. 
In particular, the non-Gaussian fluctuation measures of conserved charges such as the skewness $S\sigma$ and kurtosis $\kappa \sigma^2$ have attracted much attention recently
(see, e.g., Refs.~\cite{Stephanov1} and \cite{Stephanov2}).
The higher moments of conserved charges were suggested as probes to study 
the QCD phase structure \cite{flucQCD-1,flucQCD-2}, and have been calculated in various effective QCD models~\cite{lattice1,lattice2,QCD-1,QCD-2}. 
Experimentally, the higher moments of net-proton and net-charge multiplicity were recently measured by the STAR collaboration in Au+Au collisions in $\sqrt{s_{_{\rm NN}}} = 7.7-200$~GeV energy range~
\cite{STAR1,STAR2,STAR3}. However, no definitive conclusion regarding the existence and location of the chiral CP has been obtained yet.

In the present paper the scaled variance, skewness, and kurtosis of net-nucleon (net-baryon) number fluctuations near the critical point of nuclear matter are studied. 
A presence of the liquid-gas phase transition
in nuclear matter was reported in a large number of papers, see, e.g.,
Refs.~\cite{Finn,Pochodzalla,nm-crit}. Experimental estimates of the nuclear matter CP, $T_c\cong 17.9$~MeV and
$n_c\cong 0.06$~fm$^{-3}$, were presented recently in Ref.~\cite{nm-CP}.
At such small temperatures the effects of deconfinement and of production of new particles, such as pions, are expected to be negligible,
and the number of nucleons is essentially a conserved quantity.
Thus, very different physical pictures 
of the critical behavior in nuclear matter and near the chiral CP are evidently expected.
Nevertheless, in both cases the fluctuations of conserved charges are expected to be sensitive probes of critical behavior, and may be used to pinpoint the location of the corresponding CP.


\section{Nuclear matter with the van der Waals equation of state}

In this work we use the van der Waals (VDW) equation of state to study
the measures (\ref{osk}) of particle number fluctuations near the CP of the nuclear matter.
The VDW model contains the first-order liquid-gas  phase transition which ends at the CP.
In the canonical ensemble the classical VDW equation of state has a simple and
transparent form (see, e.g., Ref.~\cite{greiner}):
\eq{\label{eq:vdw}
p(T,n) = \frac{NT}{V-bN} - a \frac{N^2}{V^2} \equiv\frac{n\,T}{1-bn}-a\,n^2~,
}
where $V$ is the system volume, $a>0$ and $b>0$ are the VDW parameters that describe attractive and
repulsive interactions, respectively, and $n\equiv N/V$ is the particle number density.
The CP corresponds to the temperature $T_c$ and particle number density $n_c$, where
\eq{\label{p-der}
\left(\frac {\partial p}{\partial n}\right)_T = 0~,
~~~~~~\left(\frac {\partial ^2 p}{\partial n^2}\right)_T = 0~.
}

In order to apply the VDW equation of state to systems with a variable number of
particles the GCE formulation is needed (see Ref. \cite{VdW}).

In the following we consider the VDW equation of state for the system of interacting nucleons.
We restrict our consideration to  temperatures $T\le 40$~MeV,
thus, the production of new particles (like pions) is neglected.
In  addition,  both the nucleon clusters (i.e., ordinary nuclei)
and the baryonic resonances (like $N^*$ and $\Delta$) are neglected.
Within these approximations, the number of nucleons, $N$, becomes a conserved number,
and the chemical potential $\mu$ of the GCE regulates
the number density of nucleons. 

At low temperatures and/or  high particle number densities the Boltzmann
approximation becomes inadequate and
leads to unphysical negative values of the system entropy.
The generalization of the VDW equation which includes  effects of the quantum statistics was
recently proposed in Ref.~\cite{VdW-NM}. The pressure and the particle
density are then defined by the following system of two equations
for $p(T,\mu)$ and $n(T,\mu)$ functions:
\eq{
p(T,\mu) & = p^{\rm id} (T, \mu^*) - a\,n^2(T,\mu),
\label{eq:pq}\\
n(T,\mu) & = \frac{n^{\rm id}(T,\mu^*)}{1 + b \, n^{\rm id}(T,\mu^*)},
\label{eq:nq}
}
where
\begin{equation}
\label{eq:pq-2}
\mu^* = \mu~ - ~b \, p(T,\mu) - a\,b\,n^2(T,\mu) + 2 \, a \, n(T,\mu)~.
\end{equation}
The $p^{\rm id}$ and $n^{\rm id}$ are the expressions for the quantum
ideal gas pressure and particle density, respectively:
\begin{align}
p^{\rm id}(T,\mu) &= \frac{g}{6\pi^2} \int_0^{\infty} k^2dk\,
\frac{ k^2}{\sqrt{m^2+  k^2}} \nonumber \\
& \quad \times \left[ \exp\left(\frac{\sqrt{m^2+  k^2}-\mu}{T}\right)
+ \eta\right]^{-1},
\label{p-id}
\\
n^{\rm id}(T,\mu) &= \frac{g}{2\pi^2} \int_0^{\infty} k^2 dk \nonumber
\\
& \quad \times \left[ \exp\left(\frac{\sqrt{m^2+ k^2}-\mu}{T}\right) + \eta\right]^{-1},
\label{n-id}
\end{align}
where $g$ is the degeneracy factor (the number of spin and isospin states) and $m$
is the particle mass.
In Eqs.~(\ref{p-id}) and (\ref{n-id}), $\eta = +1$ for Fermi statistics, $\eta = -1$
for Bose statistics, and $\eta = 0$ for the Boltzmann approximation.

\begin{figure*}[!t]
\centering
\includegraphics[width=0.49\textwidth]{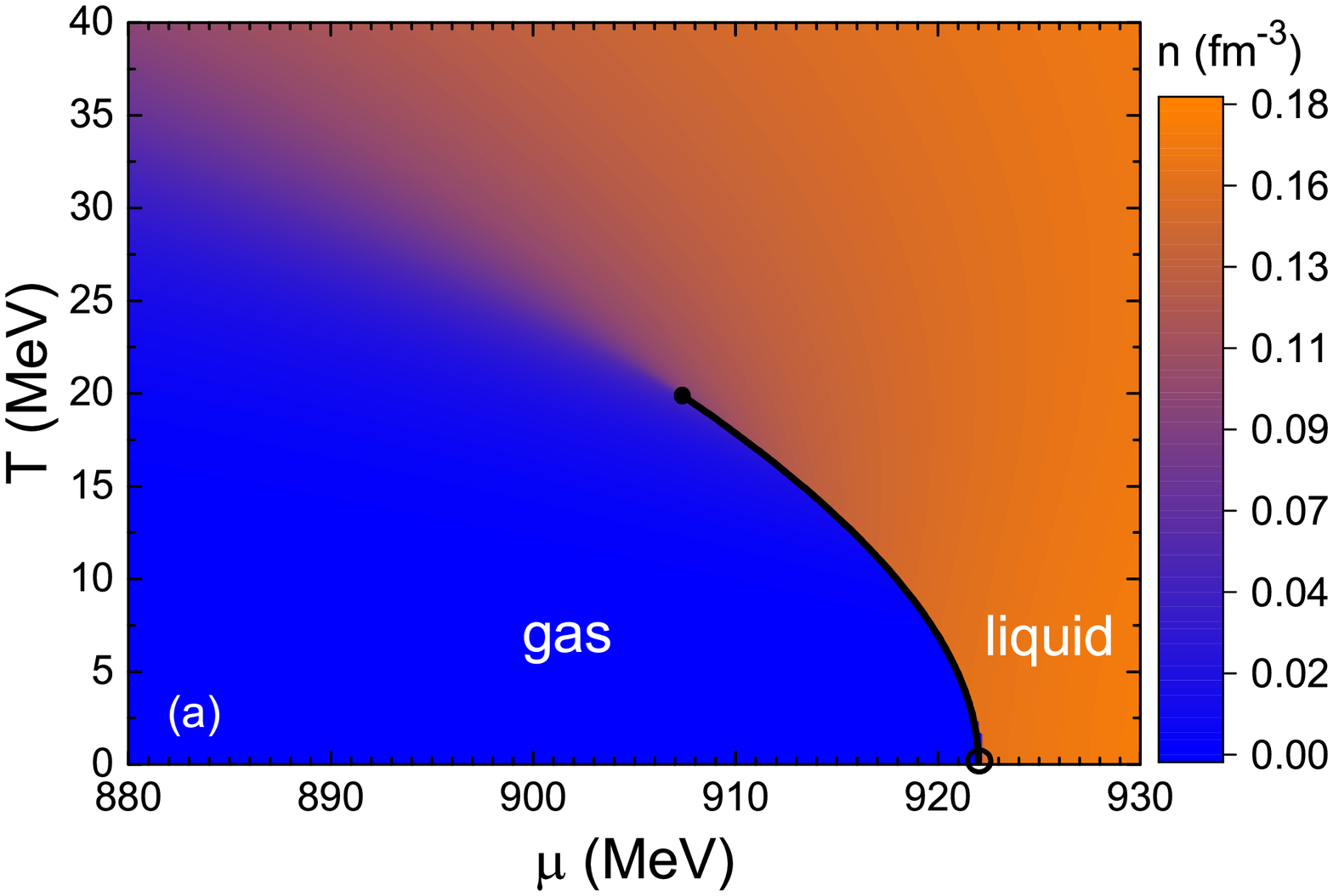}
\includegraphics[width=0.49\textwidth]{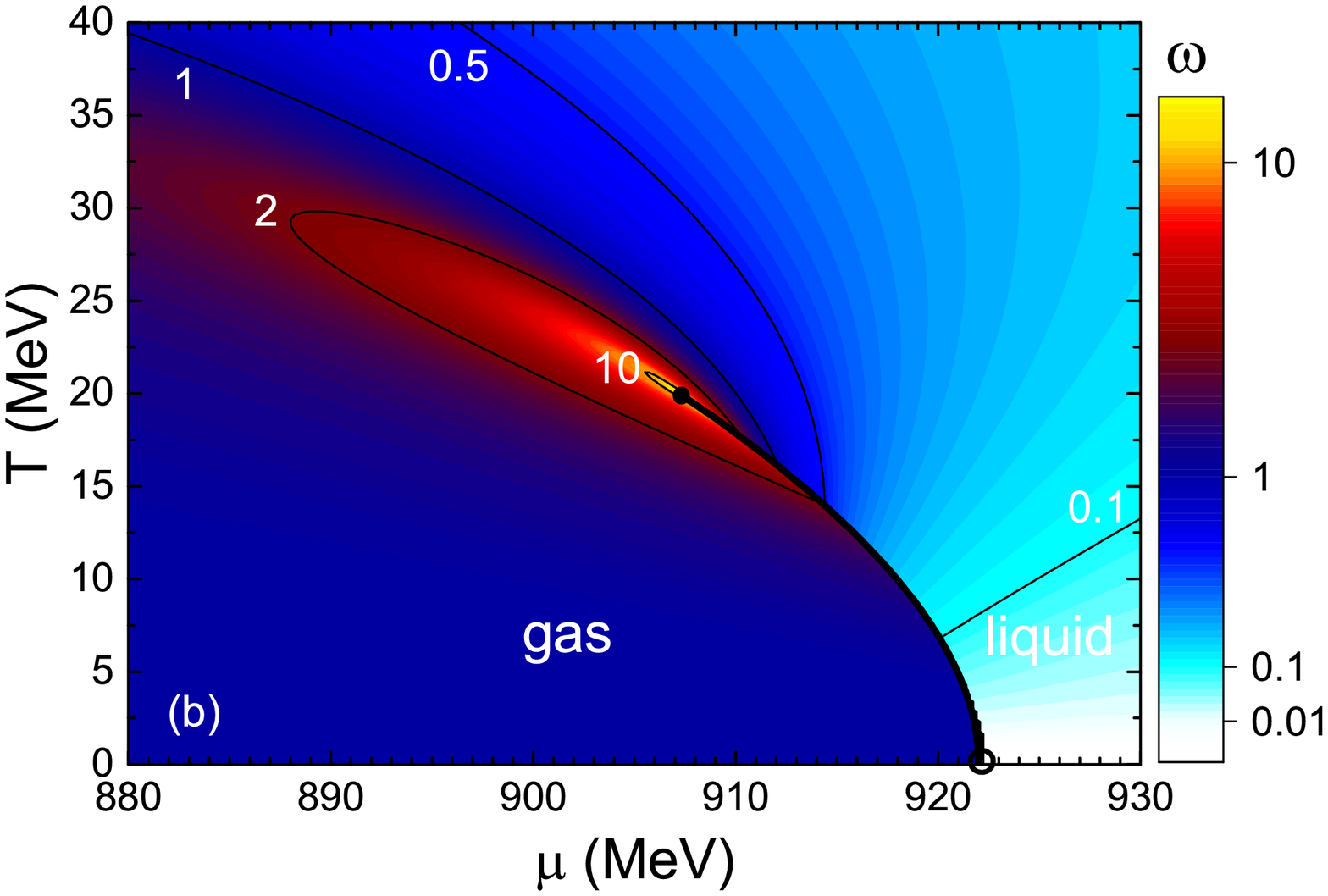}
\caption[]{
(Color online) The (a) particle number density $n(T,\mu)$ (\ref{eq:pq}) and (b) scaled variance
$\omega[N]$ (\ref{omegaN}) calculated for the symmetric nuclear matter
in $(T,\mu)$ coordinates within VDW equation of state for fermions.
The open circle at $T=0$ denotes the ground state of nuclear matter,
the solid circle at $T=T_c$ corresponds to the CP,
and the phase transition curve $\mu=\mu_{\rm mix}(T)$ is depicted by the solid  line.
}\label{fig-Tmu-n}
\end{figure*}

The VDW equation of state with Fermi statistics was used in Ref.~\cite{VdW-NM}
to describe the properties of symmetric nuclear matter.
In this case, $\eta=1$, $g=4$, and $m=938$~MeV in Eqs.~\eqref{p-id} and
\eqref{n-id}.
Parameters $a$ and $b$ are fixed to reproduce the properties of
nuclear matter in its ground state, i.e., at $T=0$ it should be
$p=0$, $n=n_0\cong 0.16$~fm$^{-3}$, and the binding energy
per nucleon $E_B=E/N - m \cong -16 $~MeV.
These conditions give
$a \cong 329$~MeV$\,$fm$^3$ and
$b \cong 3.42$~fm$^3$.
Note that particle volume parameter $b$ is connected to
its hard-core radius $r=[3b/(16\pi)]^{1/3}\cong 0.59$~fm.
In the GCE, at fixed $T$ and $\mu$, equations \eqref{eq:pq} and \eqref{eq:nq}
may have more than one solution.
In such a case a solution with the largest pressure is selected
in accordance with the Gibbs criteria (see Ref.~\cite{VdW-NM} for details).

\section{Nucleon number fluctuations near the critical point}

The phase transition line, $\mu=\mu_{\rm mix}(T)$, shown in Fig.~\ref{fig-Tmu-n}~(a),
starts from the normal nuclear matter state
with $T=0$, $\mu_0\cong 922$~MeV and ends at the CP with $T_c\cong 19.7$~MeV,
$\mu_c \cong 908$~MeV (this gives $n_c\cong 0.07$~fm$^{-3}$)
\footnote{The Boltzmann approximation $\eta=0$
leads to
$n_c=1/3b\cong 0.10$~fm$^{-3}$ and $T_c=8a/(27b)\cong 28.5$~MeV.
}.
At each point of the phase transition line,
two solutions
with different particle densities (the liquid and gas states)  and
equal pressures exist, i.e., this is a line of the first-order phase transition.
At $T>T_c$ only a single solution $n(T,\mu)$  exists.
Nevertheless, as seen from Fig.~\ref{fig-Tmu-n} (a),
a rapid although continuous change of particle number density takes place in a narrow
$T$-$\mu$ region (the so-called crossover region) even at $T>T_c$.

Using Eq.~(\ref{osk}) one calculates
the scaled variance $\omega[N]$ as
%
\eq{
\omega[N] & = \frac{k_2}{k_1} =\frac{T}{n} \, \left(\frac{\partial n}{\partial \mu}\right)_T
\nonumber
\\
& = \omega_{\rm id} (T, \mu^*) \, \left[ \frac{1}{(1-bn)^2} - \frac{2an}{T} \,
\omega_{\rm id} (T, \mu^*) \right]^{-1},
\label{omegaN}
}
where the quantity
\eq{\label{omega-id}
\omega_{\rm id} (T, \mu^*) & = 1 -
\frac{g\,\eta}{2\,\pi^2\,n} \int_0^{\infty} dk k^2
\nonumber
\\
& \quad \times\left[ \exp\left(\frac{\sqrt{m^2+k^2}-\mu^*}{T}\right) + \eta\right]^{-2},
}
with $\eta=1$ corresponds to the scaled variance of particle
number fluctuations in the ideal Fermi gas (in the Boltzmann approximation,
$\eta=0$, it is reduced to $\omega_{\rm id}=1$).

It is clearly seen from Eq.~(\ref{omegaN}) that the repulsive interactions suppress
the particle number fluctuations, whereas the attractive interactions lead to their
enhancement.
The scaled variance (\ref{omegaN}) is shown in Fig.~\ref{fig-Tmu-n} (b).
At any fixed value of temperature,  $\omega[N]\rightarrow 1$ as $\mu$ decreases.
In this case, $n\rightarrow 0$ and
$\omega_{\rm id} (T, \mu^*)\rightarrow 1$, thus, the Boltzmann ideal gas results
are recovered.
The scaled variance becomes small, $\omega[N]\ll 1$, as $\mu$ increases.
In this case, the particle number density goes to its limiting value, $n\rightarrow 1/b$.
The scaled variance (\ref{omegaN}) diverges at the CP
(note that the thermodynamic limit $V\rightarrow \infty$
is assumed). As seen from Fig.~\ref{fig-Tmu-n} (b) the large values
of $\omega[N]\gg 1$ take place along the crossover region, even far away
from the CP.

\begin{figure*}[!t]
\centering
\includegraphics[width=0.49\textwidth]{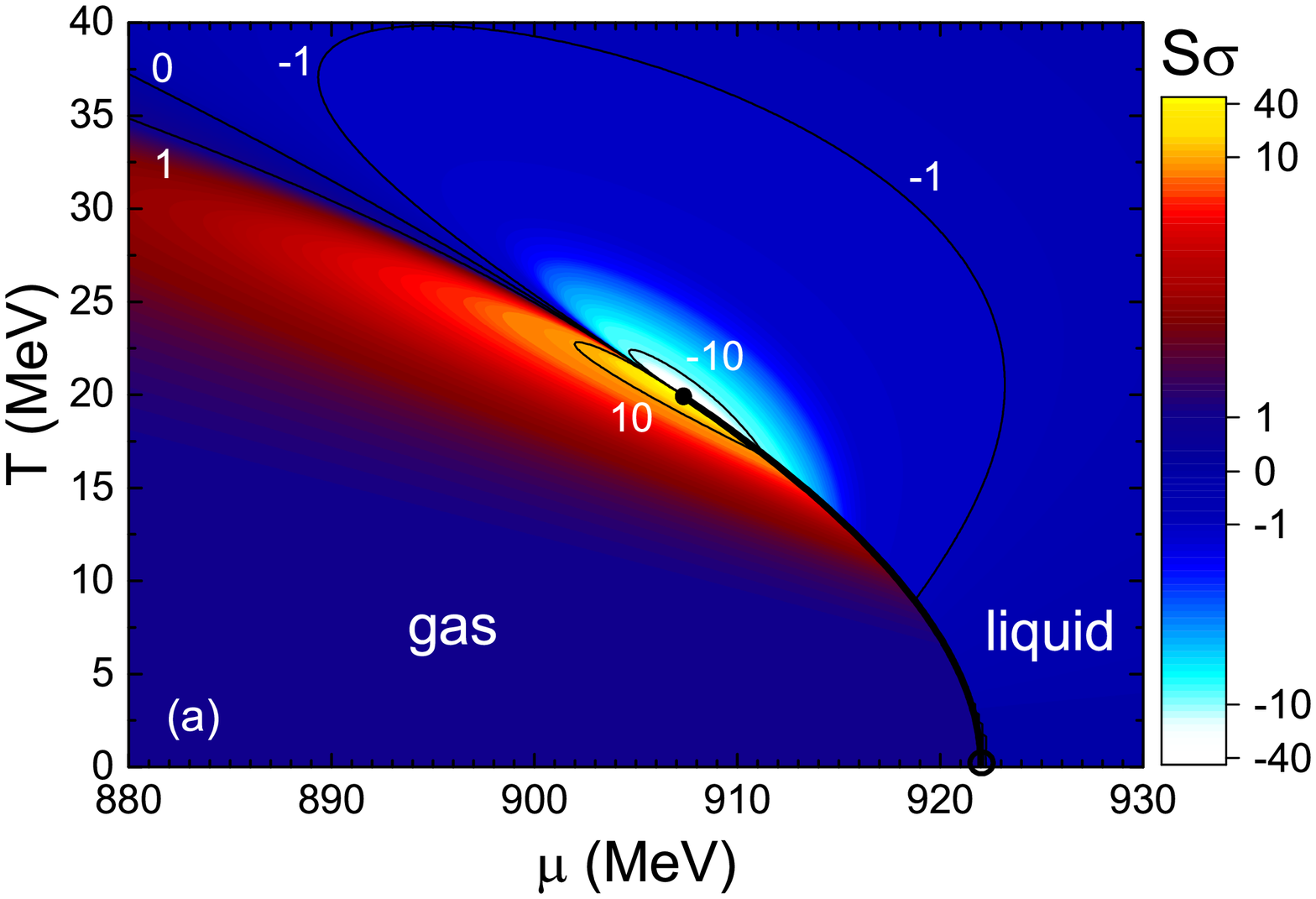}
\includegraphics[width=0.49\textwidth]{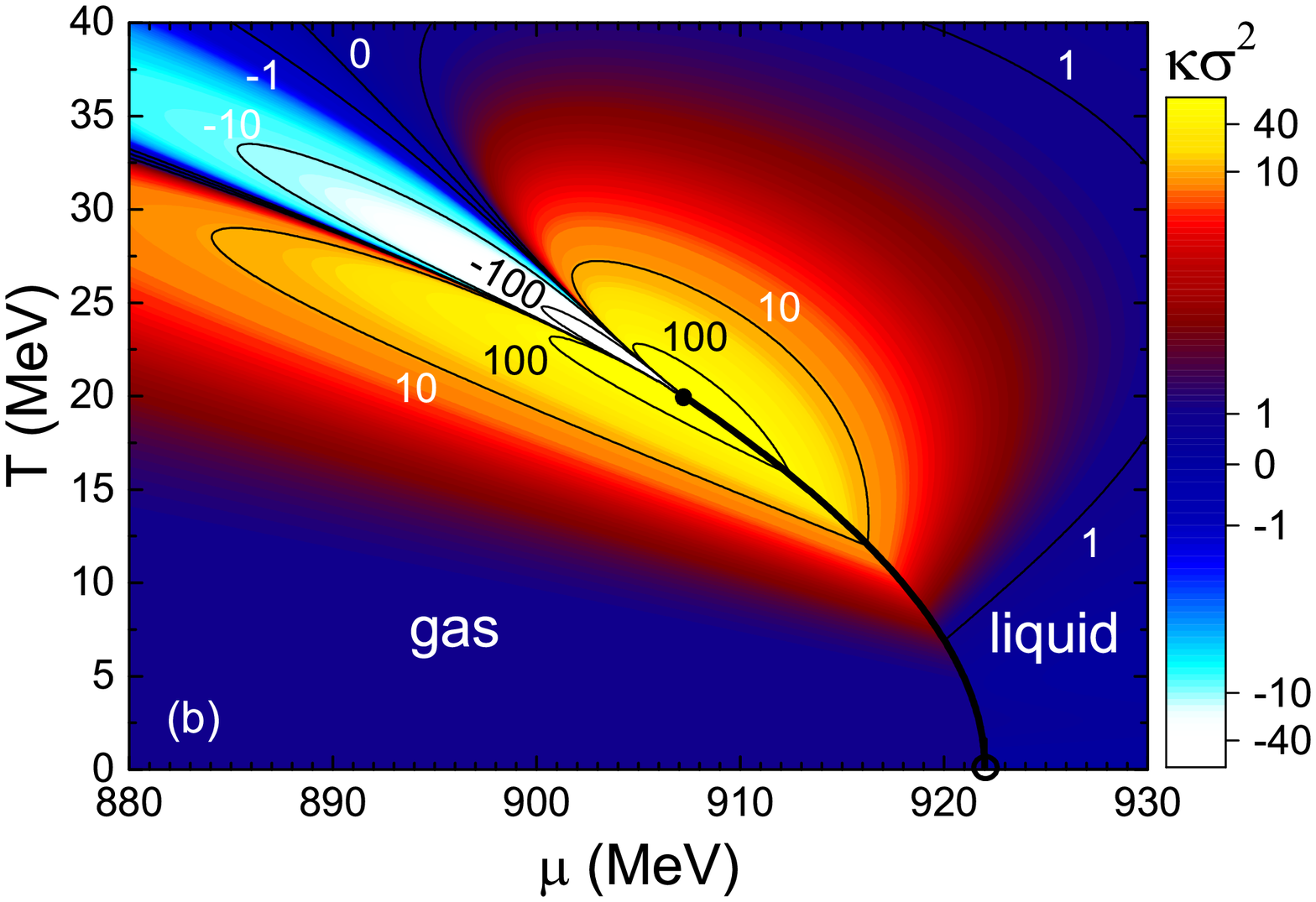}
\caption{
(Color online) The (a) skewness $S\sigma$ (\ref{eq:q-skew}) and (b)
kurtosis $\kappa\sigma^2$ (\ref{eq:q-kurt}) calculated for the symmetric nuclear matter
in $(T,\mu)$ coordinates within VDW equation of state for fermions.
} \label{Tmu-skewness}
\end{figure*}

Using Eq.~(\ref{osk}) one also calculates the skewness
\eq{\label{eq:q-skew}
S\sigma = \frac{k_3}{k_2}= \omega[N] + \frac{T}{\omega[N]} \,
\left(\frac{\partial \omega[N]}{\partial \mu}\right)_{T},
}
and the kurtosis
\eq{\label{eq:q-kurt}
\kappa \sigma^2 = \frac{k_4}{k_2}= (S\sigma)^2 + T \,
\left(\frac{\partial [S\sigma]}{\partial \mu}\right)_{T},
}
shown in Figs.~\ref{Tmu-skewness} (a) and (b), respectively.
Similarly to the scaled variance, the skewness and kurtosis  diverge
at the CP. However,
these higher moments of the particle number distribution show much richer
structures: the behavior of $S\sigma$ and $\kappa\sigma^2$  crucially depend
on the path of approach to the CP.

As seen from Fig.~\ref{Tmu-skewness} (a)
the liquid phase corresponds to $S\sigma <0$, whereas the gas phase
corresponds to $S\sigma >0$. The line of
$S\sigma =0$ goes from the critical point along
the crossover region, and the $T$-$\mu$ regions with $S\sigma\gg 1$
and $S\sigma \ll -1$ are placed just under and above this line, respectively.
From Fig.~\ref{Tmu-skewness} (b) it is also seen that the crossover region of the
phase diagram is characterized by the significantly
negative kurtosis, $\kappa\sigma^2\ll -1$.
However, outside this crossover region
one observes large positive values of the kurtosis, $\kappa\sigma^2\gg1$,
in a rather wide $T$-$\mu$ area around the CP of nuclear matter.
Qualitatively, our findings with regards to the fluctuation patterns near the CP are consistent with previous results based
on effective QCD models~(see, e.g., Refs.~\cite{flucQCD-2,QCD-1,QCD-2}), or 
on model-independent universality arguments with regards to critical behavior in the vicinity of the QCD critical point~\cite{Stephanov1,Stephanov2}.

\section{Summary}

In summary, the fluctuation signatures of the nuclear matter CP --- scaled variance, skewness, and
kurtosis --- are calculated within the quantum
formulation of the VDW equation of state. The scaled variance diverges, $\omega[N]\rightarrow \infty$,
at the CP, and the large values
of $\omega[N]\gg 1$ take place along the crossover region even far away
from the CP. The behavior of $S\sigma$ and $\kappa\sigma^2$ at the CP
is more complicated. The limiting singular values of these quantities
depend on the path of approach to the CP. The rich structures of  the skewness
and kurtosis
are observed in the wide $T$-$\mu$ area around the CP of nuclear matter.
We hope that results obtained in this paper can be useful
for the identification of the CP signatures of nuclear matter
in  heavy-ion collision experiments.

\section*{Acknowledgements}
We are thankful
to M. Ga\'zdzicki for fruitful comments and discussions.
This work was supported by the Humboldt
Foundation, by the Program of Fundamental Research
of the Department of Physics and Astronomy of the National Academy of Sciences of Ukraine,
and by HIC for FAIR within the LOEWE program of the State of Hesse.

\end{document}